\documentstyle[epsfig]{mn}

\def\simlt{\mathrel{\rlap{\lower 3pt\hbox{$\sim$}}\raise 2.0pt\hbox{$<$}}}
\def\simgt{\mathrel{\rlap{\lower 3pt\hbox{$\sim$}} \raise 2.0pt\hbox{$>$}}}
\def\lsim{\mathrel{\rlap{\lower 3pt\hbox{$\sim$}}\raise 2.0pt\hbox{$<$}}}
\def\gsim{\mathrel{\rlap{\lower 3pt\hbox{$\sim$}} \raise 2.0pt\hbox{$>$}}}

\def\Zsun{{\rm Z}_{\odot}}

\begin{document}
\title[GRBs from the early Universe]{Gamma Ray Bursts from the early Universe: predictions for present-day and future instruments}

\author[Salvaterra et al.]{R.~Salvaterra$^1$, S.~Campana$^2$, G.~Chincarini$^{1,2}$, S.~Covino$^2$, G.~Tagliaferri$^2$ \\
$1$ Dipertimento di Fisica G.~Occhialini, Universit\`a degli Studi di Milano
Bicocca, Piazza della Scienza 3, I-20126 Milano, Italy\\
$2$ INAF, Osservatorio Astronomico di Brera, via E. Bianchi 46, I-23807 Merate
(LC), Italy }

\maketitle \vspace {7cm}

\begin{abstract}
Long Gamma Ray Bursts (GRBs) constitute an important tool to study the 
Universe near and beyond the epoch of reionization. We delineate here the 
characteristics of an 'ideal' instrument for the search of GRBs at $z\ge 6-10$.
We find that the detection of these objects requires soft band 
detectors with a high sensitivity and moderately large FOV. In the light of these
results, we compare available and planned GRB missions, deriving
conservative predictions on the number of high-$z$ GRBs detectable by these 
instruments along with the maximum accessible redshift. We show that the
{\it Swift} satellite will be able to detect various GRBs at $z\ge 6$,
and likely at $z\ge 10$ if the trigger threshold is decreased by a factor of 
$\sim 2$. Furthermore, we find that INTEGRAL and GLAST are not the best tool 
to detect bursts at $z\ge 6$: the former being limited by the small FOV, and
the latter by its hard energy band and relatively low sensitivity.
Finally, future missions (SVOM, EDGE, but in particular EXIST) will provide a 
good sample of GRBs at $z\ge 6$ in a few years of operation.
\end{abstract}


\section{Introduction}

The study of the Universe at the epoch of reionization is one of the main
goal of available and future space missions. In the last few years, our
knowledge of the early Universe has been enormously increased mainly 
owing to the observation of Quasars by the SDSS survey (Fan 2006).
Long gamma ray bursts (GRB) may constitute a complementary way to study the 
cosmos and the early evolution of stars avoiding the proximity effects and 
possibly probing even larger redshifts up to $z\sim 10$. The five GRBs 
detected at $z\gsim 5$, over a sample of about
200 objects observed with the {\it Swift} satellite (Gehrels et al. 2004),
show that a large percentage of GRBs is detected at high-$z$. The current
record holder is $z = 6.29$ (Tagliaferri et al. 2005, Kawai et al. 2006).
The identification of a large 
number of GRBs at $z\ge 6$ will open a new window
in the study of the early Universe. Just to give some example, GRBs can 
be used to constrain the reionization history (Totani et al. 2006, Gallerani
et al. 2007), to study
the metallicity and dust content of normal galaxies at high-$z$ (Campana et
al. 2007), to  probe the small-scale power spectrum of density fluctuations 
(Mesinger, Perna \&  Haiman 2005). Moreover, available and future GRB missions
might be the first observatories to detect individual Population III stars,
provided that massive metal-free stars were able to trigger GRBs (see Bromm \&
Loeb 2007 for a review).
Finally, the study of GRBs at high redshift is interesting by itself.
In particular, thanks to cosmological time dilation, the study of the early 
phases of the afterglow is easier and can provide fundamental highlight on the 
central engine and burst physics.

In this paper, we delineate the main characteristics of an 'ideal' instrument 
for the search of GRBs at $z\ge 6-10$. In particular, we explore different 
observational bands, deriving the best combination of sensitivity and field of 
view (FOV) in order to detect bursts at $z\gsim 10$. In the light of these
results, we compare available and planned X-- and Gamma--ray missions, deriving
conservative predictions on the number of high-$z$ GRBs detectable by these 
instruments along with the maximum accessible redshift.

The paper is organized as follows. In Sect.~2 we briefly describe the different
models here adopted. In Sect.~3, we derive the main characteristic of an 
'ideal' instrument for exploring the high-$z$ GRB population, whereas 
predictions for available (planned) GRB missions are given in  Sect.~4 
(Sect.~5). Finally, we summarize our results in Sect.~6.

\section{Model description}

Salvaterra \& Chincarini (2007, herethereafter SC07) have computed  the luminosity function (LF) and the formation rate of long 
GRBs by fitting the observed BATSE differential peak flux number 
counts in three different scenarios:  i) GRBs follow the cosmic star 
formation and have a redshift--independent LF; ii) the GRB LF varies with 
redshift; iii) GRBs are associated with star 
formation in low--metallicity environments. 

We report here briefly the main equations used in the calculation. 
The observed photon flux, $P$, in the energy band 
$E_{\rm min}<E<E_{\rm max}$, emitted by an isotropically radiating source 
at redshift $z$ is

\begin{equation}
P=\frac{(1+z)\int^{(1+z)E_{\rm max}}_{(1+z)E_{\rm min}} S(E) dE}{4\pi d_L^2(z)},
\end{equation}

\noindent
where $S(E)$ is the differential rest--frame photon luminosity of the source, 
and $d_L(z)$ is the luminosity distance. 
To describe the typical burst spectrum we adopt the
functional form proposed by Band et al. (1993), i.e. a broken power--law
with a low--energy spectral index $\alpha$, a high--energy spectral index
$\beta$, and a break energy $E_b$, with $\alpha=-1$ and
$\beta=-2.25$ (Preece et al. 2000), and $E_b=511$ keV (Porciani \& Madau 2001).
Moreover, it is customary to define an isotropic equivalent intrinsic burst 
luminosity  in the energy band 30-2000 keV as 
$L=\int^{2000\rm{keV}}_{30\rm{keV}} E S(E)dE$. Given a normalized GRB LF, 
$\phi(L)$, the observed rate of 
bursts with peak flux between $P_1$ and $P_2$ is

\begin{eqnarray}
\frac{dN}{dt}(P_1<P<P_2)&=&\int_0^{\infty} dz \frac{dV(z)}{dz}
\frac{\Delta \Omega_s}{4\pi} \frac{\Psi_{\rm GRB}(z)}{1+z} \nonumber \\
& \times & \int^{L(P_2,z)}_{L(P_1,z)} dL^\prime \phi(L^\prime),
\end{eqnarray}

\noindent
where $dV(z)/dz=4\pi c d_L^2(z)/[H(z)(1+z)^2]$ is the comoving volume 
element\footnote{We adopted the 'concordance' model values for the
cosmological parameters: $h=0.7$, $\Omega_m=0.3$, and $\Omega_\Lambda=0.7$.},
and $H(z)=H_0 [\Omega_M (1+z)^3+\Omega_\Lambda+(1-\Omega_M-\Omega_\Lambda)(1+z)^2]^{1/2}$.
$\Delta \Omega_s$ is the solid angle covered on the sky by the survey,
and the factor $(1+z)^{-1}$ accounts for cosmological time dilation. 
Finally, $\Psi_{\rm GRB}(z)$ is the comoving burst formation rate. In this 
work, we assume that the GRB LF is described by

\begin{equation}
\phi(L) \propto \left(\frac{L}{L_{\rm cut}}\right)^{-\xi} \exp \left(-\frac{L_{\rm cut}}{L}\right).
\end{equation}

SC07 found that it is possible in all cases to 
obtain a good fit to the data by adjusting the model free parameters, i.e.
the GRB formation rate, the cut-off luminosity at $z=0$, and the
power index of the LF. Moreover, the models 
reproduce both BATSE and {\it Swift} differential counts without the
need of any change in the free parameters, showing 
that the two satellites are observing the same GRB population.
Finally, SC07 have tested the burst redshift 
distribution obtained in the different scenarios against the number of {\it
Swift} detections at $z\ge 2.5$ and $z\ge 3.5$. This procedure allows to 
constraint model results without any assumption on the redshift distribution
of bursts that have no redshift and on the effect of 
selection biases (see Fiore et al. 2007 for a detailed discussion about 
this important issue).  Models where GRBs trace the 
star formation rate (SFR) and are described by a constant LF are robustly rule 
out by available data. 
{\it Swift} detections can be explained 
assuming that the LF is evolving in redshift. In
particular, SC07 found that the typical GRB luminosity should increases with
$(1+z)^\delta$ with $\delta> 1.4$. Alternatively, the large number of 
$z\ge 2.5$ identifications may indicate that GRBs are biased tracer of the 
star formation, forming preferentially in low--metallicity environment. 
Assuming that the LF does not evolve in redshift, available data
imply a metallicity threshold for GRBs formation lower than 
$0.3\;\Zsun$ (SC07), consistently with the predictions of collapsar models 
(MacFadyen \& Woosley 1999; Izzard, Ramirez--Ruiz \& Tout 2004).

We consider here two models to compute the expected number of GRBs at 
$z\ge 6-10$ for different instruments. In the first case, we assume that
GRBs form proportionately to the global SFR and 
their typical luminosity evolves with redshift as $(1+z)^{1.4}$. 
We use here the recent determination of the SFR
obtained by Hopkins \& Beacom (2006), slightely modified to match the
observed decline of the SFR with $(1+z)^{-3.3}$ at $z\gsim 5$ suggested
by recent deep--field data (Stark et al. 2006).
This model predicts many GRBs at $z\ge 2.5$, but
is barely consistent with the number of bright GRBs at $z\ge 3.5$ 
in the {\it Swift} 2-year data sample (SC07). Moreover, this model falls 
below the lower limits on the probability to find GRBs at $z\gsim 5$ 
imposed by the five {\it Swift} confirmed detections. Since the small
number of objects does not allow to robustly rule out this model, we 
consider it as a very conservative case and we adopt the predictions of
this model as strong lower limits on the expected number of detections
at $z\ge 6-10$.

Larger, but still conservative, numbers are derived considering 
models in which GRBs form preferentially in low--metallicity envirorments.
The GRB formation rate is obtained here by convolving 
the observed SFR with the expression obtained by Langer \& Norman (2006), that
gives the fraction of galaxies at redshift $z$ with metallicity below a 
threshold metallicity $Z_{th}$.
We consider here $Z_{th}=0.1\;\Zsun$ and no luminosity 
evolution, but  similar results can be obtained for $Z_{th}=0.3\;\Zsun$
and a linear redshift evolution of the GRB LF (Salvaterra et al. 2007). These
models are found also to be consistent with the lower limit on the 
probability to detect GRBs at $z\ge 5$ (Salvaterra et al. 2007). We
will refer to this as our reference model.

We refer the interested reader to SC07 for any detail of the model 
computation and results.

\section{An ideal experiment for the search of GRB\lowercase{s} at high redshift}

\begin{figure}
\begin{center}
\centerline{\psfig{figure=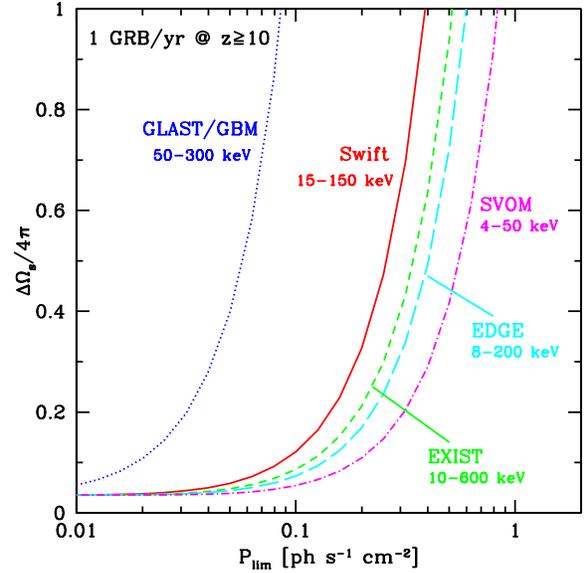,height=8cm}}
\caption{Fraction of the sky to be observed in order to detect 1 GRB per yr
at $z\ge 10$ as function of the photon flux limit of the survey.  Different
observational bands are considered: 15--150 keV (solid line; as for 
{\it Swift}); 50--300 keV (dotted line; as for GBM/GLAST); 
10--600 keV (short-dashed line; as for EXIST); 8--200 keV (long-dashed
line; as for EDGE); and 4--50 keV (dot-dashed line; as for SVOM). We adopt here our reference model.}
\label{fig:fov}
\end{center}
\end{figure}

In this Section, we explore the characteristic of an 'ideal' instrument for
the detection of GRBs at very high-$z$ (i.e., $z\sim 6-10$). 
Given the formation efficiency and
LF parameters obtained by fitting the BATSE differential peak flux 
distribution, we compute the fraction of the sky, $\Delta\Omega_s/4\pi$, 
that needs to be observed in order to detect one GRB per year at $z\ge 10$ as 
function of the peak photon flux limit of the instrument. We consider also 
different observational bands. The results obtained for our reference model 
are shown in Fig.~\ref{fig:fov}: solid line refers to 15--150 keV band (as for 
{\it Swift}), dotted line to 50--300 keV band (BATSE and GLAST/GBM), 
short-dashed line to 10--600 keV band (EXIST), and long--dashed to 8-200 keV 
(EDGE). It is clear that a hard observational band requires a very low flux 
limit in order detect GRBs at $z\ge 10$, since for high-$z$ sources this kind
of instruments sample the steep, high energy tail of the GRB spectrum. 
So, the search of bursts at very high redshift requires extremely 
sensitive detectors able to detect GRBs as faint as 0.01-0.05 ph s$^{-1}$ 
cm$^{-2}$. Note that we do not expect
any $z=10$ GRB to be present in the whole BATSE catalog.

Detectors with softer bands are more suited for detecting GRBs from  
the early Universe. Provided that the lower bound is as low as
10-15 keV, GRBs at $z\ge 10$ should trigger instruments with peak photon flux 
limits of 0.1-0.3 ph s$^{-1}$ cm$^{-2}$. We find that given a FOV, the 
sensitivity needed to observe GRBs at $z=10$ decreases by increasing the 
low-energy bound of the instrument band, whereas the detection of high-$z$ 
bursts depends only slightly on the band width. We note here that a large 
spectral coverage is still very important to derive estimates of the peak 
energy and to measure the total luminosity (energy) budget of the source.

The rapid growth of curves in Fig.~1 shows clearly that the ability to 
detect a high-$z$ GRB depends much more on the sensitivity of the 
instrument rather than on its field of view (FOV).
Soft gamma-ray detectors will detect GRBs at $z\ge 10$ even observing
only 10-15\% of the sky (1.2-2 sr), provided that their sensitivity is as
low as 0.1 ph s$^{-1}$ cm$^{-2}$. Obviously, a larger FOV implies a larger
number of detections. 

In conclusion, the observation of GRBs near or beyond the redshift of 
reionization requires soft band detectors with a high sensitivity and moderately large
FOV. An ideal instrument with a FOV of a few steradian should observe in the 
8-200 keV band (but the {\it Swift} 15--150 keV band may be enough) with 
sensitivity as low as 0.1 ph s$^{-1}$ cm$^{-2}$. Assuming 3 sr FOV, this
'ideal' instrument will be able to detect $\sim 40$ (3) GRBs at $z\ge 6$ 
($z\ge 10$) in just one year of observations.

\section{Predictions for available instruments}

In this Section we discuss the expected number of detections for 
{\it Swift}, INTEGRAL, and GLAST/GMB at $z\ge 6$ and $z\ge 10$.Model results
are given in Table~1. In the Table we report also the maximum redshift 
accessible with this instruments, $z_{max}$, corresponding to the redshift
at which $\sim 1$ GRB per year should be detected. Lower limits are obtained
with the conservative model, i.e. assuming luminosity evolution of the GRB
LF with $\delta=1.4$, whereas larger numbers are found adopting models in 
which GRBs form in low-metallicity environments. We stress here that  
also upper bounds are still conservative, being the models just consistent 
with the limits imposed by the five GRBs observed at $z\gsim 5$ (Salvaterra et 
al. 2007).

\subsection{Swift}

We compute the expected number of high-$z$ detections per year of observation
in the 15--150 keV band of {\it Swift}. We assume a FOV of 1.4 sr and a
trigger photon flux threshold of 0.4 ph s$^{-1}$ cm$^{-2}$. At this threshold,
{\it Swift} number counts are still consistent with model prediction (SC07),
indicating that the instrument observes essentially all burst above this 
photon flux. Below this limit, model prediction overestimate the number of 
{\it Swift} detection, showing that many bursts are missed by {\it Swift}.
For $P_{lim}=0.4$ ph s$^{-1}$ cm$^{-2}$, we expect $\sim 4$ GRB per year to
be detected at $z\ge 6$ in our reference model. The highest accessible redshift
is in this case $\sim 7.5$ and essentially no burst is expected to be 
detected at $z\ge 10$ during the entire {\it Swift} mission.
For our very conservative model, we expect just $\sim 1$ GRB per year to be 
detected by {\it Swift} at $z\ge 6$. We note that in this case, the 
detection  of GRB~050904 at $z=6.3$ represents an extremely rare event. 

The compilation of a large GRB sample at $z\ge 6$ (as well as many close-by 
faint GRBs) would require a lower 
trigger threshold.  Assuming to decrease the {\it Swift} threshold down to
$P_{lim}= 0.1$ (0.25) ph s$^{-1}$ cm$^{-2}$, we should be able to detect 
bursts up to $z_{max}\sim 10$ ($z_{max}\sim 8.3$) for our reference model. 
As a lower limit, we predict $z_{max}\sim 7.6$ ($z_{max}\sim 7$) for the LF 
evolution model. In both cases, 
the GRB sample at $z\ge 6$ doubles by lowering the trigger threshold from 
$0.4$ ph s$^{-1}$ cm$^{-2}$ down to $\sim 0.2$ ph 
s$^{-1}$ cm$^{-2}$. For the lowest photon flux limit here considered, we
expect to detect 3--16 GRBs per year at $z\ge 6$ and tens at $z\sim 5$. Likely,
one GRB per year should be at $z=10$. The expected redshift distribution of
GRB detections in the redshift range $z=5-10$ is shown in Fig.~2 for 
different choices of the trigger threshold.
From the plot it is clear that lowering the 
trigger threshold of {\it Swift} would largely increase the probability to
detect GRBs at high redshift, likely allowing to collect a significant sample 
of bursts at $z\ge 6$.

\begin{figure}
\begin{center}
\centerline{\psfig{figure=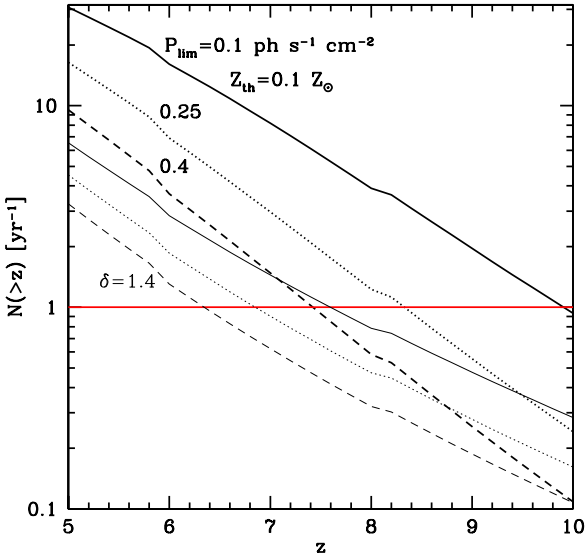,height=8cm}}
\caption{Number of GRBs per year of observation expected at redshift larger 
than $z$ for different {\it Swift} photon flux limits: 
solid line for $P_{lim}=0.1$ ph s$^{-1}$ 
cm$^{-2}$, dashed line for $P_{lim}=0.25$ ph s$^{-1}$ cm$^{-2}$ , and dotted 
line for $P_{lim}=0.4$ ph s$^{-1}$ cm$^{-2}$. Thick lines refer to our 
reference model ($Z_{th}=0.1\;\Zsun$ and no
luminosity evolution or with $Z_{th}=0.3\;\Zsun$ and $\delta=1$).  
Thin lines are obtained assuming luminosity evolution with 
$\delta=1.4$ and no metallicity threshold, and represents strong lower limits
on the number of detections. Horizontal line marks the threshold
of 1 GRB per yr. We consider the {\it Swift} 15-150 keV band and a FOV of 1.4 sr.
}
\label{fig:swift}
\end{center}
\end{figure}

Once detected, redshift measurements for GRBs at $z\gsim 5$ require rapid
follow-up observations of optical afterglow with 8-meter, ground based 
telescopes. Owing to the high competition for time on these instruments,
we need to preselect the best candidates soon after detection. This can be
done on the basis of some promptly-available information provided by {\it
Swift}, such as burst duration, photon flux, the lack of detection in the 
UVOT $V$-band, and the low Galactic extinction (Campana et al. 2007, Salvaterra
et al. 2007). Salvaterra et al. (2007) have test this procedure against the
last year {\it Swift} data, showing that this method allows to pick up
reliable $z\ge 5$ targets with an efficiency larger than 66\% and avoiding 
contamination of low-$z$ interlopers.

We note that thanks to cosmological time dilation, spectroscopical follow-up
of high redshift GRBs is easier with respect to low-$z$ ones. In fact, even a
few days after the trigger, we are still observing the afterglow in an early 
phase, when it is more luminous.

\begin{table*}\label{tab:numbers}
\begin{center}
\begin{tabular}{lcccccc}
\hline
\hline
Instrument & band & FOV & $P_{lim}$ & $z_{max}$ & GRB/yr & GRB/yr \\
 & [keV] & [sr] & [ph s$^{-1}$ cm$^{-2}$] &  & at $z\ge 6$ & at $z\ge 10$ \\
\hline
{\it Swift} & 15--150 & 1.4  & 0.4 &  6.3--7.5    & 1.3--4 & 0.09--0.1 \\
              &         &      & 0.25 &  7.0--8.3    & 2--7 & 0.16--0.25 \\
              &         &      & 0.1 &  7.5--9.9     & 3-16 & 0.3--0.9 \\
INTEGRAL/IBIS & 20--200 & 0.1 & 0.2 & 3.8--5.2 & 0.1-0.5 & $<0.01$ \\
GMB/GLAST (on-board) & 50--300 & 9 & 0.7 & 6.2--6.3    & 1.2--1.5 & $<0.1$ \\
GMB/GLAST (ground) & &  & 0.47 &    6.8--6.9    & 1.8--2.4 &  0.05--0.12 \\
\hline
SVOM & 4--50 & 2 & 1.0 & 6.7--7.4 & 2--4 & 0.1--0.13 \\ 
EDGE & 8--200 & 2.5 & 0.6 &    6.9--8     & 2--6 & 0.18--0.23 \\
EXIST & 10--600 & 5 & 0.16 &  9.7--11.3      & 11--56 & 0.9--2.8 \\
\hline
\end{tabular}
\end{center}
\caption{Characteristics of present-day and future GRB missions and predicted 
number of GRB detections per year at $z\ge 6$ and $z\ge 10$. Lower limits are 
computed by adopting luminosity evolution in the GRB LF with $\delta=1.4$, 
whereas higher, but still conservative, bounds 
are obtained for a metallicity threshold for GRB formation of 
$Z_{th}=0.1\;\Zsun$ and no luminosity  evolution (or for $Z_{th}=0.3\;\Zsun$ 
with $\delta=1$). The maximum redshift accessible, $z_{max}$ corresponds to the
redshift at which we expect $\sim 1$ GBR per year to be detected. Note that
$P_{lim}$ refers to the instrument observational band.}
\end{table*}

\subsection{INTEGRAL}

INTEGRAL is a satellite of the European Space Agency, launched on October 2002.
In almost five years, INTEGRAL detected 45 bursts, but only in a few cases 
it has been possible to measure the redshift (the current record being 
GRB~050502A at $z=3.793$).

Estimates carried out for the INTEGRAL Burst Alert System sensitivity yield
a trigger threshold of $\sim 0.14-0.22$ ph s$^{-1}$ cm$^{-2}$ for the
Imager on Board the INTEGRAL Satellite (IBIS) 
in the 20-200 keV energy band (Mereghetti et al. 2003). We find that
the probability to detect GRBs at $z\ge 6$ is comparable or even higher than
the one obtained for {\it Swift} owing to the higher sensitivity of
INTEGRAL/IBIS (see also Gorosabel et al. 2004). Unfortunately,  the above
sensitivity applies only to the central $9^\circ\times9^\circ$ of the IBIS 
where the fully instrument effective area can be used. In the external part of 
the field of view, the so called partially coded region, the sensitivity is 
worse. Even assuming $P_{lim}=0.2$ ph s$^{-1}$ cm$^{-2}$ on a FOV of 
$\sim 0.1$ sr (Gorosabel et al. 2004), only $1-3$ $z\ge 6$ GRBs should
be present in the entire INTEGRAL catalog and no one at $z\ge 10$. 
The maximum accessible redshift, computed as the redshift
at which we predict to have $\sim 1$ GRB per year, is $z\sim 5.2$ 
($z\sim 3.8$) in our reference (conservative) model.
 In conclusion, in spite of the relative good sensitivity and 
sufficiently soft observational band, we do not expect INTEGRAL to provide a 
large number of $z\ge 5$ identifications, being essentially limited by its very
small FOV.

\subsection{GLAST}

The Gamma-Ray Large Area Space Telescope (GLAST)\footnote{http://glast.gsfc.nasa.gov/} is the next NASA 
gamma-ray astrophysics mission, scheduled to be launched in January 2008
and expected to be operational for 5-10 years. It will carry two instruments:
the Large Area Telescope (LAT) and the GLAST Burst Monitor (GBM). GBM
will detect and localize burst 
monitoring $>8$ sr of the sky, including the LAT FOV.
In order to derive the expected number of bursts detectable by GLAST/GBM at
very high redshift, we consider the 50--300 keV band and a FOV of $\sim 9$ sr.
The expected on-board photon flux limit is 0.7 ph s$^{-1}$ cm$^{-2}$, whereas 
with more accurate data analysis on ground it would be possible to reach
sensitivity as low as 0.47 ph s$^{-1}$ cm$^{-2}$. As it is clear from 
Fig.~\ref{fig:fov}, GLAST/GBM is not well suited to search for bursts
at $z\ge 10$. Indeed, the hard band of the instrument together with the low 
sensitivity (with respect to the other instrument here considered) limits
the possibility to detect GRBs at redshift much larger than $z=6$. We find that
only $\sim 1$ GRB per year would be detected at $z\simeq 6.2-6.3$. No burst at
$z=10$ is expected during the entire GLAST mission.
Considering the more accurate ground data analysis, just $\sim 2$ GRBs per 
year are predicted to lie at $z\ge 6$. Although optical or near-infrared 
afterglow detection is in principle easier for high--$z$ bursts due to 
cosmological time dilation, a rapid response is still necessary in order to 
provide 
information about the first stages of the afterglow and to increase the 
probability to detect GRBs at $z\ge 6$. Thus, the information on the bursts
identified in the GLAST ground base data analysis could be available too
late for the follow--up studies. 

In conclusion, GLAST/GMB does not appear  the best instrument to search for 
GRBs at $z\ge 6$.

\section{Predictions for next generation of instruments}

The next generation of X-- and Gamma--ray instruments will improve our knowledge of
the burst physics and likely allow a systematic study of the early Universe
through the detection of a large number of GBRs. In this section, we explore
this possibility considering future missions that match the 
characteristic delineated in Section~2. Three
planned satellites will search for GRB in a soft band comparable or lower 
to that one of {\it Swift}: the Space--based multi--band astronomical
Variable Object Monitor (SVOM), the Explorer of Diffuse Emission and Gamma-ray 
burst Explosions (EDGE), and Energetic X-ray Imaging Survey Telescope (EXIST).
Expected number of $z\ge 6$ and $z\ge 10$ GRB detections for one year of
observations are given in Table~1, together with the maximum
accessible redshift.

\subsection{SVOM}

The Space--based multi--band astronomical Variable Object Monitor (SVOM)
is a Sino--French mission dedicated to the detection, localization and 
study of GRBs and other high--energy transient phenomena. The satellite launch
is scheduled in 2012. SVOM is composed
by the ECLAIRs telescope, which include the X-- and Gamma--Ray Camera (CXG,
the trigger device) and a soft X--ray Camera (ESXC), the Gamma Ray Monitor 
(GRM), the Wide Angle Cameras (WACs) and the Visible Telescope (VT). 
The ECLAIRs telescopes and the associated GRM permanently observe a large
portion of the sky (2 sr). The ECLAIRs trigger system uses the CXG data in
order to seek for the appearance of a new transient source in the gamma--ray
and determines its localization on the sky. The refined source position 
on the sky is obtained with th ESXC reaching 1 arcmin position accuracy.
Source position is then rapidly provided to the ground and also transmitted to
the VT and the WACs onboard SVOM, in order to perform visible follow--up
observation of the source. In the 4--50 keV trigger band, the expected peak 
photon flux limit for the burst detection is $\sim 1$ ph s$^{-1}$
cm$^{-2}$. As discussed in Section~3, the soft observational band of SVOM
is very suitable to search for high redshift GRBs. 
Indeed, we predict that SVOM will detect $\sim 2-4$ GRBs per year at $z\ge 6$. 
The maximum accessible redshift is $z\sim 7.4$ ($z\sim 6.7$)
in our reference (conservative) model. In spite of these good results, the 
identification of a GRB at $z\ge 10$  will be an extremely rare event: just
one burst during the entire mission is expected. In order
to detect 1 GRB per year at this limit, the SVOM trigger threshold should be 
lowered by a factor of three.

Finally, we note that, similarly to the case of {\it Swift},
GRB candidates at $z> 5$ can be selected on the basis of a few burst 
information promptly provided by ECLAIR and by the optical instruments onboard 
SVOM.

\subsection{EDGE}

EDGE\footnote{http://projects.iasf-roma.inaf.it/edge/}
 is a X--ray space  satellite, proposed in reply to the 2007 ESA Cosmic Vision
call for new space mission, that carries two X--ray telescope and has
fast re--pointing capabilities (Piro et al. 2007). The
possibility to detect GRBs is provided by a wide field monitor covering 
2.5 sr and operating in the 8--200 keV band. The expected sensitivity for 
burst detection is 0.6 ph s$^{-1}$ cm$^{-2}$ on the entire FOV here considered.
At this limit, we expect that EDGE will detect as much as 3-7 GRBs per year 
at $z\ge 6$.
EDGE should be able, at least for the most luminous bursts, to directly 
measure the GRB redshift based on intrinsic absorption in the circumbuest 
material and/or host galaxy: e.g.  in the case of high column densities with 
the metal absorption edges (and related curvature) imprint in the X--ray 
spectrum, and in case of low absorption thanks to resonant absorption lines.

The larger FOV and slightly softer observational band  with 
respect to {\it Swift} compensate in some way the higher photon flux
limit expected for EDGE.  Thus, this instrument is well suited for search 
of GRBs at $z=6$, with a few bursts expected to be detected at $z>7.5$.
In principle, for our reference model, EDGE will be able to access 
$z_{max}\sim 8$. At $z=10$, we expect only one GRB to be detected during the
three years of EDGE mission. In order to increase the number of high-$z$ 
detections, the flux limit of the instrument should be decreased by a factor 
of two. This will allow to detect $\sim 20$ GRBs per year at $z\ge 6$ and 
also a few bursts at $z=10$ (see short-dashed line in  Fig.~\ref{fig:fov}).

\subsection{EXIST}

EXIST\footnote{exist.gsfc.nasa.gov/} is a proposed hard X-ray imaging all-sky deep survey NASA mission and was 
recommended by the 2001 Report of the Decadal Survey. It is based on proven 
technology and could be launched by $\sim 2015$. One of the main goal of this 
instrument is to detect and study GRBs out to $z> 6-10$, thanks to the High
Energy Telescope (HET). HET consists of 19 wide-field hard X-ray telescopes. 
It covers the 10-600 keV band with 6 arcmin resolution (70 arcsec position
accuracy) over a 5 sr
field of view, using 6 square meters of CZT detectors. The expected 
sensitivity is as low as 0.16 ph s$^{-1}$ cm$^{-2}$ (Band 2006). The Low 
Energy Telescope (LET) complements the HET with energy coverage from 3-30 keV 
and finer spatial resolution (1 arcmin, with 10 arcsec position accuracy) 
using 1.3 square meter of silicon detectors.

Thanks to the very low trigger threshold and large FOV, EXIST will be 
the best instrument to detect high-$z$ GRBs. Almost 10-60 bursts per year
are expected to be observed at $z\ge 6$, one or three of which at $z\ge 10$.
Even for the very conservative model, EXIST will be able to detect GRBs up
to $z\sim 10$ and to collect a significant number of bursts at $z\ge 6$ during
the entire mission.

\section{Conclusions}

We have explored the characteristics of an ideal mission for the search of 
GRBs near and beyond the epoch of reionization, i.e. $z> 6-10$. 
In particular, we considered different observational bands, deriving the best 
combination of sensitivity and 
field of view in order to detect bursts at $z\gsim 10$. We found that such an 
experiment requires soft band detectors and high sensitivity, whereas large FOV
or wide energy coverage are less important. Assuming 3 sr FOV, an observational
band of 8-200 keV, and a sensitivity as low as 0.1 ph s$^{-1}$ cm$^{-2}$,
this instrument would be able to detect $\sim 40$ (3) GRBs at $z\ge 6$ 
($z\ge 10$) in one year of mission.

In the light of these results, we compared available and planned GRB missions, 
deriving conservative predictions on the observable number of GRBs at $z\ge 6$
and $z\ge 10$ along with the maximum accessible redshift. We have shown
that {\it Swift} is a viable tool to detect GRBs at $z\sim 6$. At
the actual trigger threshold, $1.3-4$ GRBs per year should be identified
above this redshift. We discuss also the possibility of increasing the number 
of high-$z$ detections by lowering the {\it Swift} trigger threshold. We 
found that the number of detectable GRBs doubles by lowering this by about 
50\%. Assuming to be able to further lowering the trigger threshold down to 
0.1 ph s$^{-1}$ cm$^{-2}$, {\it Swift} should detect $\sim 1$ GRB per year 
at $z\ge 10$. 

The INTEGRAL and GLAST satellites do not appear the best tools to search for 
GRBs at very high redshift. The former is limited by the very small FOV whereas
the GLAST hard energy observational band and relatively low sensitivity do
not allow to detect more than 1 GRB per year at $z\ge 6$. No GRB at $z\ge 10$ 
is expected during the entire mission of both instruments. 

Finally, we show that future missions, like SVOM, EDGE, and in particular 
EXIST, will be able
to collect a good number of GRBs at $z\ge 6$ in a few years of operations. This
sample can be use to study the early Universe, possible providing 
strong constrain on the reionization process (Gallerani et al. 2007), 
and deriving estimate on the
star formation and metallicity/dust content in normal high-$z$ galaxies.

\section*{Acknowledgments}

We are grateful to L. Piro and  L. Amati for providing information about the
EDGE mission, N. Omodei for discussion about GLAST, and S.~Mereghetti for 
providing the sensitivity curve expected for SVOM.

\end{document}